\documentclass[aps,prl,reprint,showpacs,superscriptaddress]{revtex4-1}

\usepackage{amsmath}
\usepackage{graphicx}
\usepackage{CJK}

\def\vGp{{{\bf G}_\parallel}}
\def\vb{{\bf b}}
\def\vr{{\bf r}}
\def\vrp{{\vr_\parallel}}

\begin{document}

\newcommand{\kin}{{\bf k}}
\newcommand{\kinz}{k_z}
\newcommand{\kinpar}{{\bf k}_\parallel}

\newcommand{\kfree}{{\bf q}}
\newcommand{\kfreepar}{{\bf q}_\parallel}
\newcommand{\kfreez}{q_z}

\begin{CJK*}{UTF8}{bsmi}
\title{First-principles theory of low-energy electron diffraction and quantum interference in few-layer graphene}

\author{John F. McClain}
\affiliation{Integrated Applied Mathematics Program, University of New Hampshire, Durham, New Hampshire 03824, USA}

\author{Jiebing Sun}
\affiliation{Department of Physics, Michigan State University, East Lansing, Michigan 48824, USA}

\author{Karsten Pohl}
\affiliation{Department of Physics and Materials Science Program, University of New Hampshire, Durham, New Hampshire 03824, USA}
\author{Jian-Ming Tang
(湯健銘)
}
\email{Corresponding author: jmtang@mailaps.org}
\affiliation{Department of Physics and Materials Science Program, University of New Hampshire, Durham, New Hampshire 03824, USA}

\begin{abstract}

We present a computationally efficient method to incorporate density-functional theory into the calculation of reflectivity in low-energy electron microscopy. The reflectivity is determined by matching plane waves representing the electron beams to the Kohn-Sham wave functions calculated for a finite slab in a supercell. We show that the observed quantum interference effects in the reflectivity spectra of a few layers of graphene on a substrate can be reproduced well by the calculations using a moderate slab thickness.

\end{abstract}


\maketitle 
\end{CJK*}

Low-energy electron microscopy (LEEM) is a powerful microscopy tool for {\it in-situ} surface analysis which combines the precision control of high-energy electron beams and the surface sensitivity of low-energy electrons \cite{Tromp.2000,Bauer.2007,Altman.2010}. LEEM often utilizes specularly reflected electrons (bright-field imaging) in the low-energy range ($\lesssim 30$ eV), a region not accessible to conventional low-energy electron diffraction (LEED). Strong quantum interference effects can be observed in this energy range, e.g., recent LEEM observations \cite{hibino_microscopic_2008,sutter_epitaxial_2008,Riedl.Coletti.2009,locatelli_corrugation_2010} of few-layer graphene (FLG) show interesting layer-dependent oscillations in the reflectivity spectra for energies less than 10 eV.

Theoretical analysis of the LEED spectra is traditionally carried out using multiple scattering theory with spherically symmetric (muffin-tin) potentials \cite{Pendry.1974,VanHove.Weinberg.1986}. For directionally bonded materials, such as semiconductors, the self-consistent electronic potentials in the interstitial region (outside the muffin-tin spheres) can be significant, particularly for the very low-energy electrons in LEEM analysis \cite{Rubner.Kottcke.1995}. In addition, the surface structures used in these calculations are determined by fitting the calculated diffraction spectra to the experimental data. Therefore, the development of a first-principles theory for LEEM is highly desirable.

\begin{figure}
\includegraphics[width=\columnwidth]{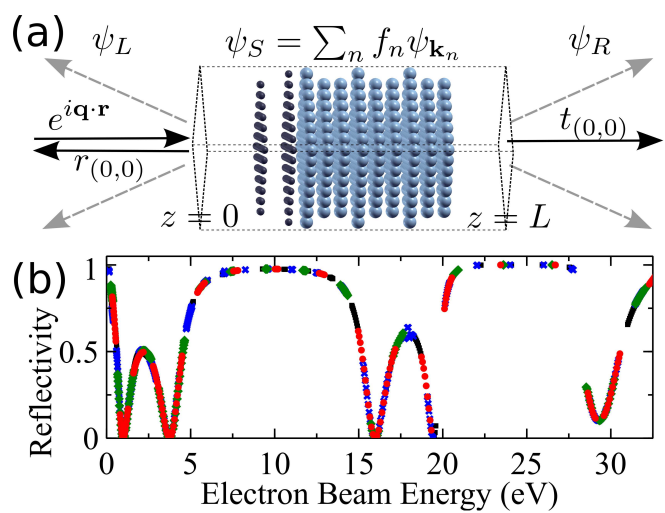}
\caption{(color online) (a) Schematic showing the wave matching configuration. Black (solid) arrows represent plane waves involved in the specular energy regime. Grayed (dashed) arrows represent higher-order diffraction beams that are relevant at higher energies. (b) Reflectivity of 3 layers of free-standing graphene, calculated using 4 different supercell sizes: 9 (red circle), 10 (green diamond), 14 (blue cross), and 15 (black square) in-plane lattice constants.}
\label{Scattering_diagram}
\end{figure}

There are two common approaches to studying the scattering of electrons off a surface. One is to solve for a semi-infinite crystal \cite{Krasovskii.Schattke.1997,*Krasovskii.Schattke.1999}, and the other is to solve for a finite slab. A hybrid approach has also been proposed recently \cite{Krasovskii.2004}. Although the semi-infinite setup may seem to be a logical starting point, it is not a simple generalization of the bulk problem because the surface states do not exist in a bulk crystal. One needs to introduce the concept of complex band structures and deal with the numerical instabilities associated with exponentially growing solutions. It also becomes computationally demanding if one attempts to solve for the ground state self-consistently. As a result, progress has been limited to simple systems. On the other hand, using a finite slab surrounded by a vacuum region in a supercell has become a popular choice, despite the drawback of finite-size effects. Advancements in computation speed have made it possible to carry out density-functional (DFT) calculations in large supercells. For LEEM, we believe that a wave-matching scheme utilizing the slab geometry is more efficient because scattering states need not be part of the self-consistency process and the electron mean-free paths are not very long in the energy range of interest \cite{Zangwill.1988}. Another practical reason for choosing the slab setup is that the scheme can be easily integrated into many well-developed general-purpose DFT packages. We have recently become aware that a very similar approach with a different implementation of the matching has been used to study free-standing FLG \cite{Feenstra.Srivastava.2013} and FLG on metal substrates \cite{srivastava_feenstra_2013}. Our matching scheme requires a smaller vacuum region and can include multiple diffracted beams.

We will first outline our wave-matching scheme, and then show applications of the method to FLG systems with and without a substrate. Our calculated results for free-standing FLG exhibit oscillations in reflectivity for energies between 0 and 6 eV, in good agreement with the experimental LEEM spectra of FLG observed on various substrates. The actual positions of the peaks and valleys can vary with the substrate. The number of oscillations is correlated to the number of graphene layers, a fact often used to determine the number of graphene layers. There is a second set of oscillations in the 14--21 eV range, which is less pronounced in the experimental observations, presumably due to damping and short electron coherence lengths in that energy range. There is also a single valley in all spectra around 30 eV. We will discuss the correspondence between these features and the bulk graphite band structure later in this paper. To study the effects due to the substrates, we calculate FLG on Ni(111)-(1x1) and find that the FLG features dominate those of the bare Ni(111) when graphene layers are added, as seen in experiments. Our results for FLG on Ni also show that the valleys in the LEEM spectra due to graphene appear only with more than one graphene layer, consistent with our results for free-standing FLG.

\begin{figure}
\includegraphics[width=\columnwidth]{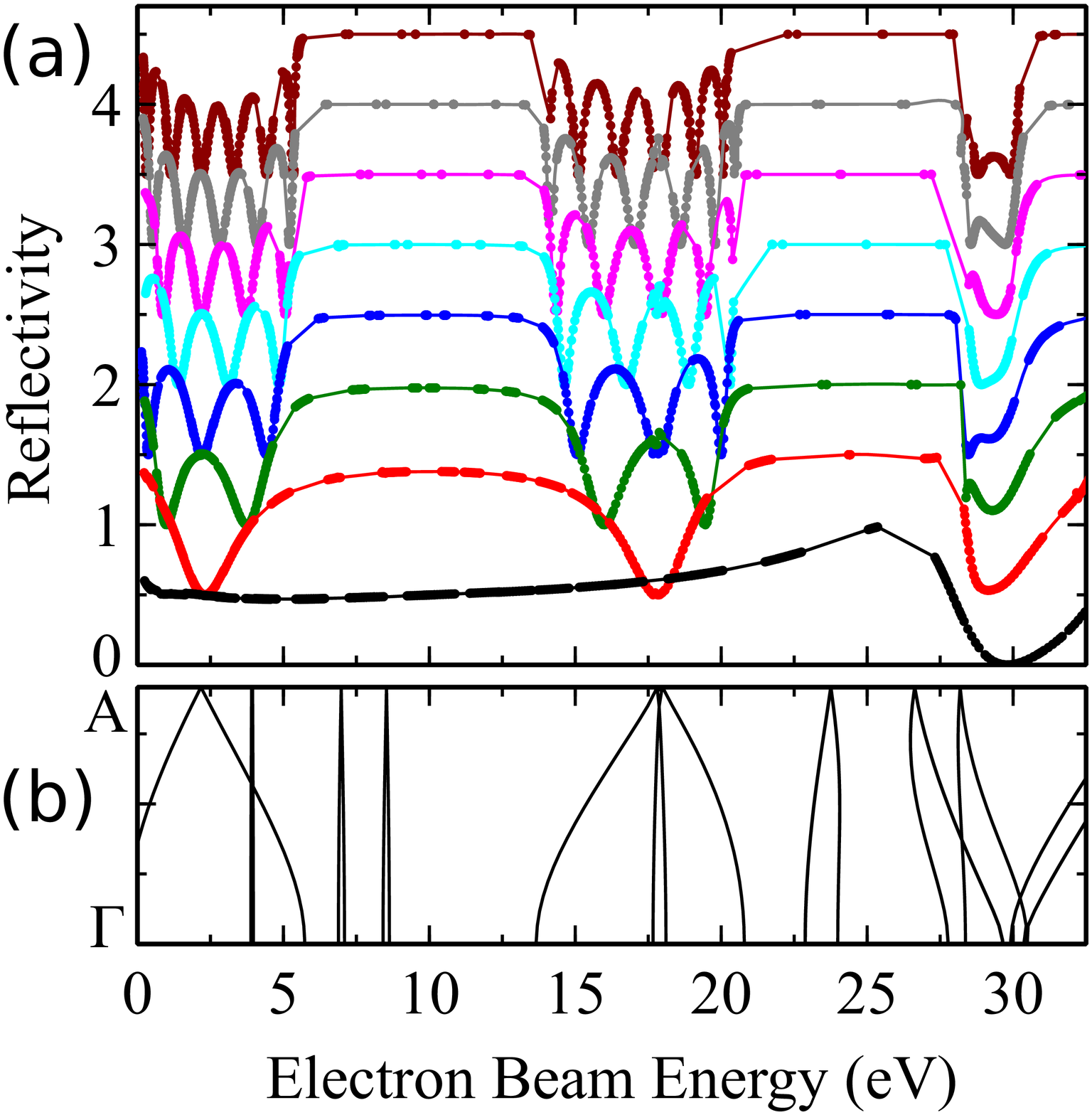}
\caption{(color online) (a) Reflectivity spectra for 1 to 8 layers of free-standing graphene, bottom to top. The calculated data points are shifted vertically in half integers for clarity. The lines are a guide to the eyes. (b) Band structure of bulk graphite along the hexagonal axis from $\Gamma$ to $A$. The Fermi level of the bulk graphite is aligned with the Fermi level of the 8-layer free-standing graphene.}
\label{Free_FLG}
\end{figure}

The schematic of our scattering setup is shown in Fig.~\ref{Scattering_diagram}(a). The central region is treated as a supercell in self-consistent DFT calculations and is composed of vacuum regions surrounding a slab.
The surface normal is the $-\hat z$ direction and there is one incident wave from $z<0$ with a wave vector $\kfree=(\kfreepar,\kfreez)$ and energy $E$. The number of reflected and transmitted waves depends on the energy and the incident angle of the electron beam. 
According to Bloch's theorem, an eigenstate inside the supercell with the given energy $E$ can be written as
\begin{eqnarray}
\psi_{\kin}(\vr) & = & e^{i\kin\cdot\vr}\sum_{\vGp}c_{\kin}(\vGp, z)e^{i\vGp\cdot\vrp} \;,
\end{eqnarray}
where $\vGp$ are the in-plane reciprocal lattice vectors, and the coefficients $c_\kin(\vGp,z)$ are calculated in terms of Fourier components for the supercell system, $d_{\kin}(\vGp,G_z)$, via
\begin{eqnarray}
c_\kin(\vGp,z) & = & \sum_{G_z} d_{\kin}(\vGp, G_z) e^{iG_z z} \;,
\end{eqnarray}
where $G_z$ are the reciprocal lattice points along $\kinz$. A wave function $\psi_S(\vr)$ with energy $E$ inside the supercell can be constructed as a linear combination of these energy eigenstates,
\begin{eqnarray}
\psi_S({\bf r}) & = & \sum_{n=1 }^D f_n \psi_{\kin_n}(\vr) \;,
\end{eqnarray}
where $D$ is the degeneracy at this energy. The periodicity of the potential for the system leads to the Bragg condition, $\kfreepar = \kinpar + \vGp$.
So the general solution outside the supercell can be written as 
\begin{eqnarray}
\psi_L(\vr) & = & e^{i\kfree\cdot\vr}+\sum_{\vGp} r_{(\alpha,\beta)} e^{i(\kfreepar+\vGp)\cdot\vrp-iq_z'z} , \\
\psi_R(\vr) & = & \sum_{\vGp} t_{(\alpha,\beta)} e^{i(\kfreepar+\vGp)\cdot\vrp+iq_z'z},
\end{eqnarray}
where $r_{(\alpha,\beta)}$ and $t_{(\alpha,\beta)}$ are the amplitudes of reflection and transmission, respectively, for each $\vGp=\alpha\vb_1+\beta\vb_2$ in the primitive basis, $\{\vb_1,\vb_2\}$, and
\begin{eqnarray}
q_z' & = & \sqrt{q_z^2-|\vGp|^2-2\kfreepar\cdot\vGp} \;.
\end{eqnarray}
For each $\vGp$, enforcing the continuity of the wave function and its first derivative at each supercell boundary gives two constraints on these amplitudes.

In the following, we restrict our discussions to normal incidence, $\kfreepar=\kinpar={\bf 0}$, and specular reflection only, i.e., the energy is less than $E_c=\hbar^2 b_s^2/2m$, where $b_s=\min(|\vb_1|,|\vb_2|)$ and $m$ is the free electron mass. With these restrictions, we find the generic case to be $D=2$ (a pair of solutions with Bloch vectors $\kin_1$, $\kin_2$ with $k_{2,z}=-k_{1,z}$). The matching problem can then be reduced to an inhomogeneous system of 4 equations for exactly 4 unknowns, involving only components corresponding to $\vGp=(0,0)$ and
\begin{eqnarray}
\begin{pmatrix}
1 \\
q_z \\
0 \\
0 \\
\end{pmatrix}
& = &
\begin{pmatrix}
-1 & \phi_{\kin_1}(0) & \phi_{\kin_2}(0) & 0 \\
q_z & \phi'_{\kin_1}(0) & \phi'_{\kin_2}(0) & 0 \\
0 & \phi_{\kin_1}(L) & \phi_{\kin_2}(L) & -e^{iq_zL} \\
0 & \phi'_{\kin_1}(L) & \phi'_{\kin_2}(L) & -q_ze^{iq_zL}
\end{pmatrix}
\begin{pmatrix}
r_{(0,0)} \\
f_1 \\
f_2 \\
t_{(0,0)}
\end{pmatrix}
\label{eq:matching}
\end{eqnarray}
where $\phi_{\kin_j}(z)= e^{i\kin_j\cdot{\bf r}}c_{\kin_j}({\bf 0},z)$, and $\phi_{\kin_j}'(z)$ is the first derivative.
The reflectivity is then computed as $|r_{(0,0)}|^2$. 

In exceptional cases, the number of linearly independent solutions inside the supercell does not match the number of constraints. The artificial periodicity in the $z$ direction creates artificial band gaps, and these gaps lead to voids in the spectra. At the $\Gamma$-point and at the perpendicular Brillouin zone boundaries, only one solution is found at each energy, so these points are excluded.  However, most of these voids can be filled by carrying out matching with different vacuum thicknesses as shown in Fig.~\ref{Scattering_diagram}(b).  We also find bands (of supercell solutions) with two-fold degeneracy, leading to too many solutions and, thus, too many unknowns, but, for the systems we studied, these solutions do not pose a problem, as they are removed from the matching procedure due to their localization in the $z$ direction, an issue described below.

Further, we find that, independent of matching constraint considerations, certain solutions are unfit for this matching procedure.  Some solutions, lying on bands with limited dispersion, have wavefunctions that are bound states in the $z$ direction (i.e., with exponential decay in the vacuum). These states cannot interact with scattering states and, so, must be removed from the matching procedure. Other solutions, lying near the intersections between bands with limited dispersion and bands with substantial dispersion, create problems in the matching without appearing so clearly unfit for scattering. To filter out these states, we use the determinant of the matrix in Eq.~(\ref{eq:matching}) as a guide.  Feenstra et al. and Srivastava et al. seem to have encountered this issue as well and dealt with it using a measure of the degree to which the solution is plane-wave-like near the supercell boundary. (See \cite{srivastava_feenstra_2013} and Supplemental Material to \cite{Feenstra.Srivastava.2013}.) Both approaches require arbitrary thresholds to be set.  We believe that a more rigorous approach is ultimately desirable.

DFT calculations are performed using the open-source {\sc Quantum ESPRESSO} software package \cite{Giannozzi.Baroni.2009}.
The ground state and the self-consistent potential are first calculated using a $14\times 14\times 1$ Monkhorst-Pack grid. For free-standing FLG, we use the experimental lattice parameters of graphite: 2.46 \AA\ for the in-plane lattice constant and 3.36 \AA\ for the spacing between graphene sheets. For FLG-Ni systems, we fix the bottom Ni layers to the calculated lattice constant, 3.42 \AA , of bulk Ni and relax the interlayer spacing of the graphene layers and the top 3 layers of Ni. The spectra presented here are calculated using two different supercell sizes, sufficient to resolve the relevant features and to check convergence. The minimum distance from the surface layers of the slab to the supercell boundary is 8 \AA. For consistency between different supercell sizes, the ground-state calculation for each system in the larger supercell is performed with the structure determined for that system in the smaller supercell. For the scattering states, we use 18 pairs of degenerate solutions, equally spaced along $\kinz$ in the Brillouin zone. The local density approximation, ultrasoft pseudopotentials, and Perdew-Zunger exchange-correlation are used in all calculations. Gaussian smearing of 0.001 Ry is employed in free-standing FLG and 0.01 Ry in FLG-Ni. Dipole corrections \cite{Bengtsson.1999} are added to the Ni side of the vacuum region to account for the asymmetry of the FLG-Ni systems. The plane wave cutoff energy is 50 Ry.

Figure \ref{Free_FLG}(a) shows our calculated reflectivity versus electron beam energy for one to eight layers of graphene. The zero energy is referenced to the self-consistent potential in the vacuum regions. The cutoff energy $E_c$ for the first-order diffraction is about 33 eV. The results show that for one layer of graphene, the reflectivity has no apparent feature until the minimum around 30 eV. This minimum is due to the quantum interference within a single layer and is observed in thicker FLG due to the fact that they are weakly-coupled layered systems.

For two or more layers, two sets of oscillations appear in the spectra, between 0 and 6 eV and between 14 and 21 eV. The number of minima increases with the number of layers of FLG. The positions of the two sets of oscillations correspond well to the two dispersive bands in bulk graphite \cite{hibino_microscopic_2008}. Oscillations in the lower energy range are observed  in the measured LEEM spectra for FLG up to 9 layers on SiC \cite{Hibino.Kageshima.2008}. For the higher energy range, only the first minimum corresponding to a bilayer is observed in FLG with two or more layers. This may be explained by the stronger damping in this energy range \cite{Strocov.Blaha.2000,*barrett_elastic_2005}, and it is consistent with the fact that the typical electron mean-free path at 20 eV is about 10 \AA\ \cite{Zangwill.1988}. Note that we have yet to include damping in our modeling. For energies between these three major valleys, nearly complete reflection is found, even for just two layers. These significant levels of reflection can be correlated to gaps or nearly flat bands in the bulk graphite band structure.

\begin{figure}
\includegraphics[width=\columnwidth]{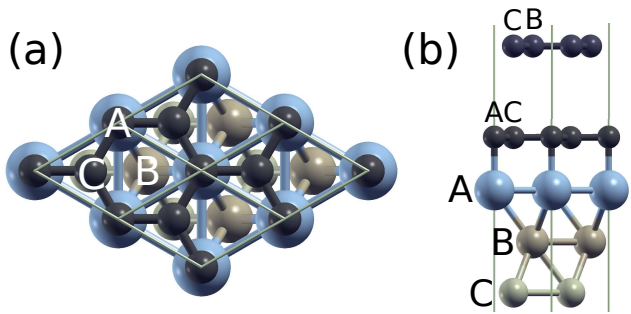}
\caption{(color online) Structure models of graphene on Ni(111) used in DFT calculations. (a) Top view along the surface normal direction of single-layer graphene on Ni. The hexagons of the graphene layer are centered on the Ni atoms (B sites) in the second layer.
(b) Side view of 2 layers of graphene on Ni with the hexagons of the top graphene layer centered on the Ni atoms (A sites) in the first layer.
}
\label{Interface_Struct}
\end{figure}

\begin{figure}
\includegraphics[width=\columnwidth]{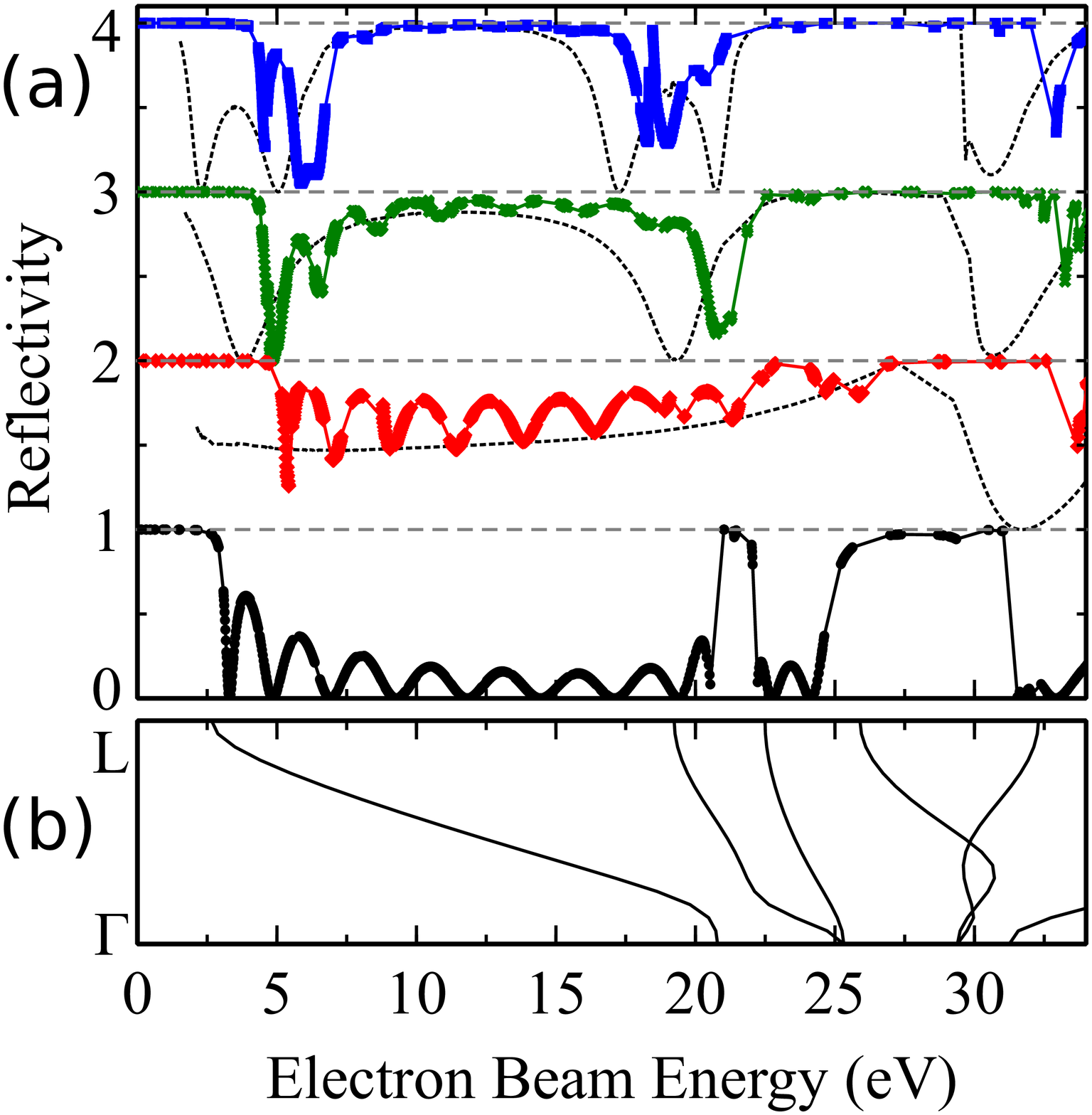}
\caption{(color online) (a) Reflectivity spectra for 0--3 layers of graphene on 10 layers of Ni, bottom to top. The calculated data points are shifted vertically in integers for clarity. The solid lines are a guide to the eyes. The free-standing FLG results with the same number of graphene layers are shown in dotted lines for comparison. The dotted lines have been shifted in energy by the dipole correction. (b) Band structure of bulk Ni along (111) from $\Gamma$ to $L$. The Fermi level of bulk Ni is aligned with the Fermi level of the 10-layer Ni slab.}
	\label{FLG_on_Ni}
\end{figure}

So far, we have shown that calculations of free-standing FLG with no damping and no fitting parameters already give very good agreement with experimental results of FLG on different substrates. To understand the influence of the substrate and illustrate the finite-size effects of a slab, we study FLG on a Ni(111) substrate. We find that the FLG related features remain qualitatively the same.

The FLG-Ni system admits a $1\times 1$ interface lattice structure \cite{gamo_atomic_struct_1997,sun_graphene_on_Ni_2010,lahiri_graphene_growth_2011} because the lattice constant of graphene is shorter than the in-plane lattice constant of Ni(111) by only $1\%$. For the first layer of graphene, the two carbon atoms in a unit cell are located above the Ni atoms in the first and the third layers, as shown in Fig.~\ref{Interface_Struct}(a). For Bernal stacking, the second layer of graphene has two possible placements. We use the model given in Ref.~\onlinecite{sun_graphene_on_Ni_2010}, as shown in Fig.~\ref{Interface_Struct}(b). We calculate the spacing between Ni and graphene to be $\sim 2.05$ \AA. Relaxation produces little change in the spacing between Ni layers from the value in the bulk. The spacing between the graphene layers in a bilayer system is 3.18 \AA. For 3 layers of graphene on Ni, the spacings are 3.27 and 3.17 \AA\ from top to bottom.

Figure \ref{FLG_on_Ni} shows the calculated reflectivity spectra for slabs consisting of 10 layers of Ni covered by 0--3 layers of graphene. We choose the vacuum potential on the FLG side to be zero. The potential near the Ni surface is higher by approximately 1.9, 1.5, and 1.3 eV for one, two, and three layers of graphene, respectively. For the bare Ni surface, there are two apparent peaks between 20--30 eV, consistent with experimental observations \cite{sun_graphene_on_Ni_2010}. The valley below 20 eV corresponds well to a dispersive band in bulk Ni. The reflectivity in the valley oscillates in a fashion similar to that of the free-standing FLG results. The number of oscillations is related to the number of layers in the Ni slab. In the presence of a large number of layers and damping, these oscillations should reduce to a smooth envelope. Our results are quite consistent with the results calculated for a semi-infinite Ni surface \cite{flege_Ni_oxidation_2011}. For the system covered by one layer of graphene, the reflection curve does not have significant features, except for the oscillations due to the finite slab of Ni. For two and three layers, the two valleys in the reflectivity spectra stand out quite clearly, and the amplitudes of the finite-slab oscillations reduce significantly for energies between the valleys. This is consistent with the high reflectivity of FLG in these energy ranges. The number of oscillations in each valley is tied to the number of graphene layers, in the same way as for free-standing FLG. It also appears that the valleys are shifted to higher energies, in agreement with experimental observation \cite{Odahara.Ishikawa.2011}.

In conclusion, we have developed an efficient wave-matching approach to calculate LEEM reflectivity spectra based on DFT.
This method can be directly integrated into many general-purpose solid-state DFT packages. 
Our results for FLG systems and for a clean Ni surface show very good agreement with experimental observations.
We show that the reflectivity spectra of FLG-Ni systems are dominated by FLG-related features even when damping is not included in our calculations. The Ni substrate sharpens the valley features and shifts the positions of the minima to higher energies.

The authors acknowledge James B. Hannon's early work on this approach and thank him for his thoughtful discussions. This work is supported by NSF DMR-1006863.

\bibliography{leem}

\end{document}